\documentclass[conference]{IEEEtran}
\IEEEoverridecommandlockouts
\usepackage{cite}
\usepackage{amsmath,amssymb,amsfonts}
\usepackage{algorithmicx}
\usepackage[algo2e]{algorithm2e}
\usepackage{algorithm}
\usepackage{graphicx}
\usepackage{siunitx}
\usepackage{textcomp}
\usepackage{xcolor}
\usepackage{soul}
\usepackage{siunitx}
\usepackage{comment}
\usepackage{epstopdf}

\usepackage{url}
\usepackage[left=1.62cm,right=1.62cm,top=1.778cm]{geometry}
\usepackage{balance}
\def\BibTeX{{\rm B\kern-.05em{\sc i\kern-.025em b}\kern-.08em
    T\kern-.1667em\lower.7ex\hbox{E}\kern-.125emX}}
\begin{document}

\title{On Transfer Learning for  a Fully Convolutional Deep Neural SIMO Receiver \\
\thanks{This research is supported by the HORIZON JU-SNS-2022-STREAM-B-01-02 CENTRIC project (Grant Agreement No.101096379).}
}

\author{

\IEEEauthorblockN{Uyoata E. Uyoata, Ramoni O. Adeogun}

\IEEEauthorblockA{Wireless Communication Networks Section, Department of Electronic Systems, Aalborg University, Denmark\\ Email: \{ueu,ra\}@es.aau.dk}


}

\maketitle

\begin{abstract}
Deep learning has been used to tackle problems in wireless communication including signal detection, channel estimation, traffic prediction, and demapping. Achieving reasonable results with deep learning typically requires large datasets which may be difficult to obtain for every scenario/configuration in wireless communication. Transfer learning (TL) solves this problem by leveraging knowledge and experience gained from one scenario or configuration to adapt a system to a different scenario using smaller dataset. TL has been studied for various stand-alone parts of the radio receiver where individual receiver components, for example, the channel estimator are replaced by a neural network. There has however been no work on TL for receivers where the entire receiver chain is replaced by a neural network. This paper fills this gap by studying the performance of fine-tuning based  transfer learning techniques for various configuration mismatch cases using a deep neural single-input-multiple-output(SIMO) receiver. Simulation results show that overall, partial fine-tuning better closes the performance gap between zero target dataset and sufficient target dataset.
\end{abstract}


\section{Introduction}
Deep learning has been applied to power control, packet scheduling, congestion control, and resource allocation. Physical layer tasks formerly performed by handcrafted mathematically heavy approaches now have deep learning equivalents that provide comparable results. For example in \cite{8052521, 9153885}, some deep learning architectures were used for physical layer tasks including channel estimation, signal detection, and demapping. 
To achieve reasonable results using deep learning, large datasets are needed to enable a deep neural network extract sufficient features from the input samples for tuning the parameters of the network. Depending on the size of designed deep neural network, high computing resources are also needed to train them in considerable time. Traditionally, machine learning including deep learning is task and domain specific, meaning that for each new task or domain, a new dataset needs to be acquired and new training cycles need to be scheduled as well. This is particularly challenging for wireless communication where designs are naturally prone to varying propagation conditions, different constellations and/or waveforms, and user mobility. 


Techniques exist to reduce training sample size and leverage knowledge gained from already available models trained on large datasets. Transfer learning is one such technique. 
In transfer learning, a target model builds on the experience gained from a pretrained model to perform a target task translating to a significant reduction in the amount of data required for updating the weights of the target model. It is worth mentioning that TL relies on the assumption that there exists a relationship between the source and target domain/task. 


Transfer learning has been employed for various wireless communication tasks, see e.g., \cite{9868088, 10122171, 8805017, 9388873, 9175003 } and the references therein. These papers have shown that TL techniques have the potential to reduce the amount of data required for training without compromising the achieved performance. For example, the work in \cite{9868088} has shown that fine-tuning the weights of some layers of the target model can improve the block error rate (BLER) performance of a neural network-based channel estimator. 
The work in \cite{10122171} has also shown that a target model with a small dataset can leverage the knowledge gained from a source model trained for channel estimation in a NOMA-MIMO scenario to offer a performance comparable to training the target model from scratch using a large dataset. Similar observations were also made in the paper for various system configurations including channel model, power allocation, and modulation. Transfer learning for intra-cell and inter-cell channel quality indicator (CQI) prediction  is the focus in \cite{8805017}. To that end, both convolutional neural network (CNN) and Long Short Term Memory (LSTM) networks wherein selected transferred layers are fine-tuned using target data were used. The results show that fine-tuning the LSTM model performs better than the convolutional neural network model when target domain data is available and vice versa. Fine-tuning is also used in \cite{9388873} for transferring models between cell sites in two different countries. 
In \cite{9175003}, meta-transfer learning is shown to enable the use of a small dataset at the target model for downlink channel state information (CSI) prediction.

In general, there seems to be an increasing interest in transfer learning for physical layer communication tasks resulting in a number of published works within the last few years. Most of the existing research contributions have however focused on specific receiver components, for example, the channel estimator  and not the complete receiver chain. This motivates our study of the performance of transfer learning techniques for a deep neural receiver \cite{DeepRx, E2ERex} that encapsulates the various components of a communication receiver. Specifically the objective of this paper is to study the transferability of a deep neural receiver using finetuning based transfer learning techniques for various configuration mismatches. The SIMO receiver \cite{Siona} considered in this paper performs channel estimation, equalization and demapping.  The contributions of this paper include: 
\begin{enumerate}
    \item the investigation of the transferability of a single-input-multiple-output (SIMO) deep neural receiver similar to the one presented in  \cite{Siona}. 
    \item presentation of the application of feature extraction and fine-tuning techniques for adapting the SIMO deep neural receiver to various configuration mismatches. 
    \item presentation of results of extensive simulations for various source model - target model mismatch scenarios such as channel model, modulation schemes, and subcarrier spacing (SCS) mismatches. 
\end{enumerate}

\section{End-to-End Receiver}
\begin{figure}
 \centering
 \includegraphics[width=0.8\linewidth]{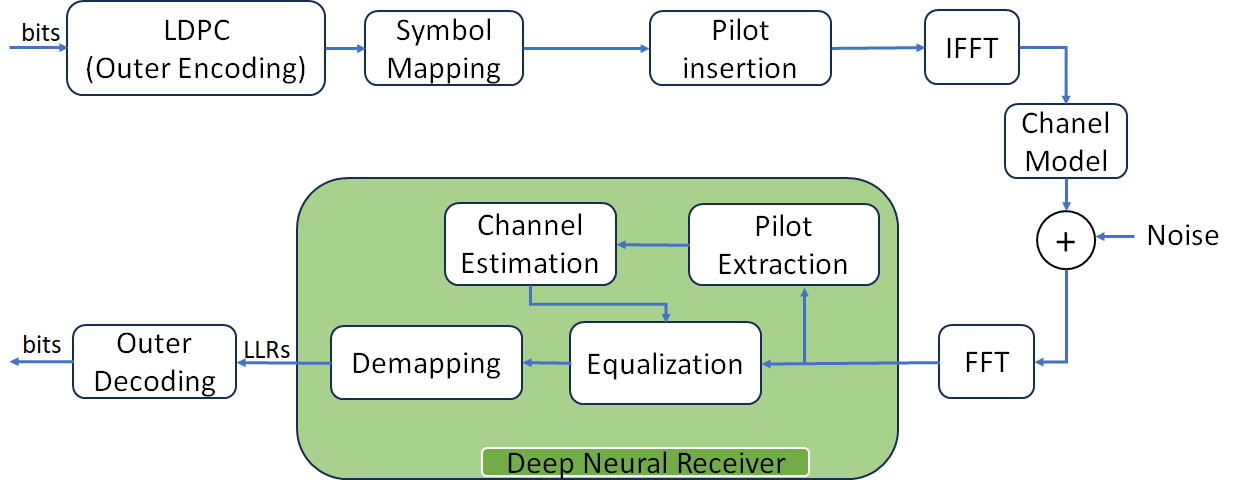}
 \caption{Illustration of Sionna link level simulation used for data generation and receiver benchmarking \cite{Siona}}
 \label{fig_0}
 \end{figure}

This work considers an end-to-end communication chain where uniformly distributed bits randomly generated at the transmitting end are first encoded using a Low Density Parity Check (LDPC) code. Then the code words are mapped to a constellation, and the resulting symbols are placed on physical resource blocks.  Inverse Fast Fourier Transform converts the symbols into an Orthogonal Frequency Division Multiplexing (OFDM) waveform having 14 symbols within a transmission time interval for transmission through the wireless channel. 
Pilots are inserted at specified indices in the resource grid, corresponding to indices 2 and 11 in the time domain. More so the first 5 and the last 6 subcarriers are used as guard bands. 3GPP compliant channel models are used for the wireless channel and these models are CDL A, B, C, D, E and UMi channel models \cite{3gpp} . An illustration of this end-to-end system is shown in Figure \ref{fig_0}.
  \begin{figure}
 \centering
 \includegraphics[width=\linewidth]{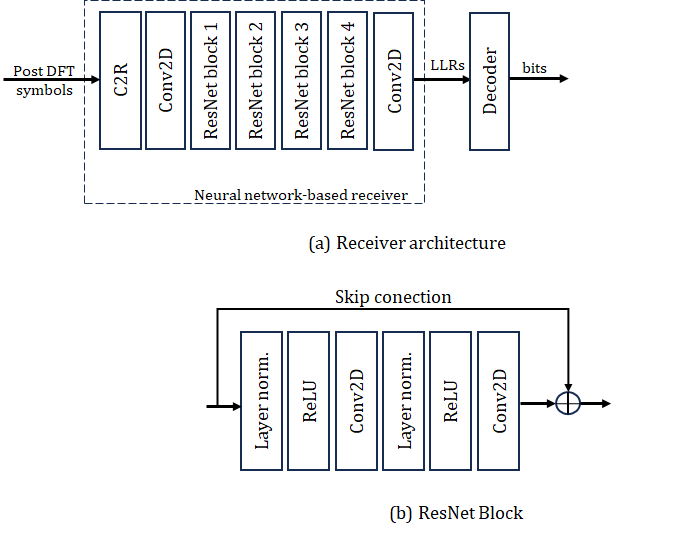}
 \caption{Neural network architecture}
 \label{fig_01}
 \end{figure}
At the receiving end,  Additive White Gaussian Noise (AWGN) is added to the waveform which goes through Discrete Fourier Transformation (DFT) before being fed into the deep neural receiver. 

In the deep neural receiver, four ResNet blocks sit between two Conv2D blocks as Figure \ref{fig_01} illustrates.
 Each ResNet block is composed of two Conv2D blocks, two linear normalization (norm.) blocks, and two Rectified Linear Units (ReLUs). Moreover, there is a skip connection between the input and output of each ResNet block. Details of the neural network layers are given in Table \ref{T_0}.
\begin{table}
\caption{Details of neural network architecture}
\centering 
\begin{tabular}{|l|l|l|l|} 
\hline 
Layer & Channels & Kernel size& Dilation rate\\ [0.5ex]
\hline 
Input Conv2D & 128 &(3,3)& (1,1)\\
\hline
ResNet 1 & 256 &(3,3)& (1,1)\\
\hline
ResNet 2 & 256 &(3,3)& (1,1)\\
\hline
ResNet 3 & 256 &(3,3)& (1,1)\\
\hline
ResNet 4 & 256 &(3,3)& (1,1)\\
\hline
Output Conv2D & number of &(3,3)& (1,1)\\
 & bits/symbol  &&\\
\hline
\end{tabular}
\label{T_0}
\end{table}

 The input convolutional layer of the deep neural receiver is fed resource grids of dimension defined by the number of OFDM symbols ($F$) and the number of subcarriers ($S$). Specifically the dimension of the input to the deep neural receiver is ($M \times N_{rx} \times F \times S$) where $M $ is the batch size and $N_{rx}$ is the number of receive antennas. The symbols on these resource grids are complex and so undergo a pre-processing that separates real and imaginary parts of the symbols and then stacks them. The output of the  deep neural receiver are log likelihood ratios (LLRs) with which the binary cross entropy (BCE) is determined over all: bits/symbol, OFDM symbols, subcarriers, and batches. The training objective of the deep neural receiver is to minimize the loss function (bit metric decoding rate)  defined as :
 \begin{equation}
L = 1 - \dfrac{1}{MSFK} \sum_{m = 1}^{M}\sum_{s = 1}^{S}\sum_{f = 1}^{F} \sum_{k = 1}^{K} BCE(B_{m,s,f,k}, LLR_{m,s,f,k}),
 \end{equation}
 where M is the training batch size which is the number of samples processed by the neural receiver before internal parameter updates, S is the number of OFDM subcarriers, F is the number of OFDM symbols, K is the number of bits in each symbol,  $B_{m,s,f,k}$ is the $k_{th}$ coded bit transmitted on the (s,f) resource element. For the $k_{th}$ coded bit transmitted on the $(s,f)_{th}$ resource element for the $b_{th}$ batch example, the logit determined by the deep neural receiver is given by $LLR_{m,f,s,k}$.
Finally, an LDPC decoder is applied to the output of the deep neural receiver.  
 
Training a deep neural receiver can be epoch based using a static dataset or can be performed by generating training symbols on the fly. The latter approach is used in this paper and the details of the end-to-end receiver setup is available at \cite{Siona}. The deep neural network architecture used in this paper is similar to the architecture used in \cite{9508784} and has been employed in end-to-end learning with pilot-less communication.

\RestyleAlgo{ruled}
\SetKwComment{Comment}{/* }{ */}


\begin{figure}
\centering
\includegraphics[width=0.8\linewidth]{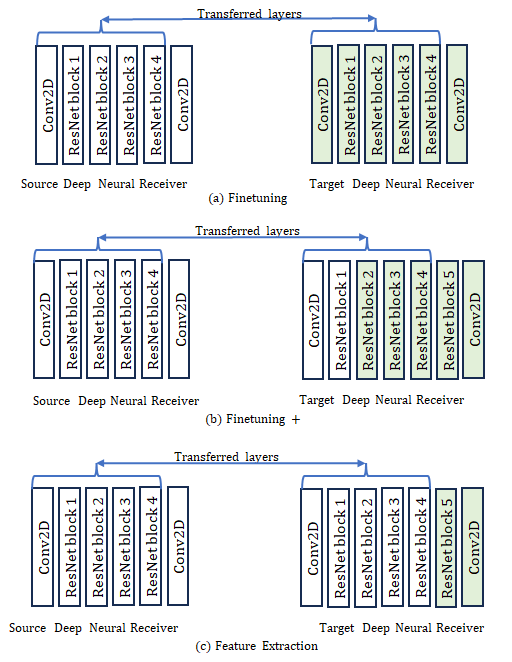}
\caption{Transfer learning techniques (the green shaded blocks are modified with target dataset)}
\label{fig_app}
\end{figure}
\section{Transfer learning Techniques}
Based on the definition in \cite{5288526}, transfer learning is defined thus: given a source domain $(D_S)$, a target domain $(D_T)$, source task $(T_S)$ and target task $(T_T)$, transfer learning aims to employ the knowledge obtained from $D_S$ and $T_S$ to better the outcome of the predictive function $f_{T}(.)$ in $D_T$ provided $D_{S} \neq D_{T}$ or $T_{S} \neq T_{T}$.
 The domain, $D$ is composed of a feature space ($X$) and a marginal probability distribution $P(X)$. For this paper, both the source task and the target task are the same. And this task is post-DFT symbol decoding.  However, the source domain and the target domain are not the same being that the feature space in the source and target domains are not derived from the same configuration. Basically the source dataset and the target datasets are not the same. So the problem addressed in this paper qualifies as a transfer learning problem.

For this problem, the transfer learning techniques evaluated in this paper are based on fine-tuning and feature extraction. In fine-tuning, the weights from the source neural network ($\Phi_S$) are partially or completely modified by target dataset ($X_T$).

We consider two fine-tuning based techniques namely:
\subsubsection{Fine Tuning} where  complete source model weights ($\Phi_S$) modification with target dataset ($X_T$) is performed, and \subsubsection{Fine Tuning +} where partial source model weights ($\Phi_S$) modification with target dataset ($X_T$) is performed . Here, the architecture of the source model is also modified by adding one more ResNet block.

Conversely, in \textit{Feature Extraction}, none of the weights transferred from the source model ($\Phi_S$) is modified by target dataset ($X_T$), but  only the weights of new layers are modified by target dataset.
\RestyleAlgo{ruled}
\SetKwComment{Comment}{/* }{ */}
 A picture depiction of these studied techniques is given in Fig. \ref{fig_app} where the green filled rectangles indicate layers modified by target dataset. To avoid cluttering the figure, input to the neural network, interconnection between the layers of the neural network and the output from the neural network are left out of Figure \ref{fig_app}.  The logical flow describing these techniques are shown in Algorithms \ref{alg2} and \ref{alg3}.  
 
 Although there is some similarity between the architecture of \textit{Fine Tuning +} and \textit{Feature Extraction}, in \textit{Feature Extraction} the features of the source model are left intact whereas \textit{Fine Tuning +} modifies the features of some layers of the source model and the features of the added layer. 
 
There is no constraint on the selection of the particular layer beyond which fine-tuning is performed. For the  \textit{ Fine Tuning} case, all the weights of the source model transferred to the target model are modified using target dataset and since the source and target model architecture are the same, it can be implied that all layers of the target model are finetuned. For the \textit{Feature Extraction} case, all the weights of the source model transferred to the target model are frozen. The weights of the last two layers of the target model are modified by the target dataset. Finally, for \textit{Fine Tuning +} case, only the weights of  the first two layers of the target model are frozen whereas the weights of the rest of the layers are modified by target dataset. 
Other hyper parameter configurations used in this paper including  the number of layers and the learning rate are illustrated in Figure \ref{fig_app} and listed in Table \ref{T} respectively. The number of layers for both the \textit{Feature Extraction} and the \textit{Fine Tuning +} cases are fixed to 7(that is, 1 input Conv2D layer, 5 ResNet Layers and 1 output Conv2D layer). For the \textit{Fine Tuning} case, the number of layers is 6(that is, 1 input Conv2D layer, 4 ResNet Layers and 1 output Conv2D layer). The learning rate is fixed at 0.001 for all considered techniques.

\begin{algorithm}[!ht]
\caption{$\longrightarrow$ \textbf{Technique 1 and 2: Fine tuning}}
\label{alg2}
\textbf{Input:} Source deep neural receiver ($f_{\Phi_{S}}$), $k$, $N_{iterations}$\\
\textbf{Output:} BLER of target deep neural receiver ($f_{\Phi_{T}}$)  \\
\eIf{"Fine Tuning" }
{
Retain $f_{\Phi_{S}}$ model architecture\\
 Load source network parameters $\Phi_{T} \longleftarrow \Phi_{S}$\\
    \For{i = 1 to  $N_{iterations}$} {
Generate target dataset, $X_{T}$\\
Input dataset into target network \\
Update the parameters of $f_{\Phi_{T}}$ for trainable layers 
    }

}
{
\If{"Fine Tuning +"}{
Add an extra ResNet block\\
Load source network parameters $\Phi_{T} \longleftarrow \Phi_{S}$\\
 Freeze $k$ layers of $f_{\Phi_{T}}$ \\
\For{i = 1 to  $N_{iterations}$} {
Generate target dataset, $X_{T}$\\
 Input dataset into target network \\
 Update the parameters of $f_{\Phi_{T}}$ for trainable layers 
}
}
}
 Instantiate target model $f_{\Phi_{T}}$ and load weights, $\Phi_{T}$\\
 Evaluate target model on test data\\
 Output BLER  of target model or receiver
\end{algorithm}

\begin{algorithm}
\caption{$\longrightarrow$ \textbf{Technique 3: Feature extraction}}
 \label{alg3}
 \textbf{Input:} Source deep neural receiver ($f_{\Phi_{S}}$), $N_{iterations}$\\
 \textbf{Output:} BLER of target deep  neural receiver ($f_{\Phi_{T}}$)  \\
 Add an extra ResNet block\\
 Load source network parameters $\Phi_{T} \longleftarrow \Phi_{S}$\\
Freeze all layers of $f_{\Phi_{S}}$ \\
\For{i = 1 to  $N_{iterations}$} 
{
Generate target dataset, $X_{T}$ \\
Input dataset into target network \\
 Update the parameters of $f_{\Phi_{T}}$ for trainable layers
}

 Instantiate target model $f_{\Phi_{T}}$ and load weights, $\Phi_{T}$ \\
 Evaluate target model on test data \\
Output BLER  of target model or receiver   
\end{algorithm}

\section{Evaluation}
In this section, benchmark approaches are explained, and simulation results are discussed. The data generation, training of the deep neural receiver, and evaluation of the techniques are performed using Sionna, a Python based open source library for physical layer simulations  \cite{Siona}. Simulation parameters listed in Table~\ref{T} are some of the configurations used in this paper. Note that for this work, a sample is an OFDM resource grid having 14 symbols and 128 subcarriers.
\begin{table}
\caption{Simulation Parameters} 
\centering 
\begin{tabular}{|l|l|} 
\hline 
\textbf{Parameter} & \textbf{Value}\\ [0.5ex]
\hline 
Resource Grid & \num{14} symbols\\
&\num{128} subcarriers\\
&Guard carriers index : [$5,6$]\\
& subcarrier spacing: $30$, $60$ and \SI{120}KHz\\
&  Pilots symbols index: [$2$ $11$]\\
\hline
Channel Models & 3GPP UMi, CDL –A, B, C, D and E\\

\hline
Modulation & QPSK, $16$QAM, $64$QAM\\
 Coding& LDPC \\
\hline
Deep Neural Receiver &Batch size, $M$: $128$ \\

& Optimizer: Adam \\
& Input shape - ($M\times N_{rx} \times F \times S$)\\
& Output shape - ($M \times n_{bits}$)\\
\hline
Transmitter (Tx) & $1$ Tx, $1$ Tx Antenna\\
 configuration & \\
\hline
Receiver (Rx) & $1$ Rx, $1$ $\times$ $1$ antenna array,\\
 configuration& dual polarization\\
\hline
Direction & Uplink \\
\hline
Training & Source dataset size: $3,480,000$   \\
& Target dataset size: $348,000$ \\
&Signal-to-noise-ratio(SNR) range: -$4$ - \SI{8} dB\\
&Learning rate: $10^{-3}$\\
\hline
\end{tabular}
\label{T}
\end{table}
Techniques used as benchmarks to compare the performance of the transfer learning techniques explained in the previous section are:

\noindent\textbf{1. Without TL:$|X_T|=\alpha|X_S|$}: where the target deep neural receiver or target model is trained on a target dataset ($X_T$) having a size $|X_T|=\alpha|X_S|$,  $(0 < \alpha\leq 1)$, and evaluated on a test dataset derived from $D_T$.

\noindent\textbf{2. Model transfer: $X_S$ $\rightarrow$ $X_T$}: where the deep neural receiver is trained on the source dataset, $X_S$ but evaluated on   target dataset, $X_T$.

As stated  earlier, we consider three kinds of configuration mismatch between the source and target models  viz: modulation, channel model, and sub-carrier spacing mismatches. The performance evaluation results for each of these are presented in the sequel. For this paper, since receiver performance is studied, BLock Error Rate (BLER) is used as the performance metric.  

 \begin{figure}[!t]
 \centering
  \includegraphics[width=1.0\linewidth]{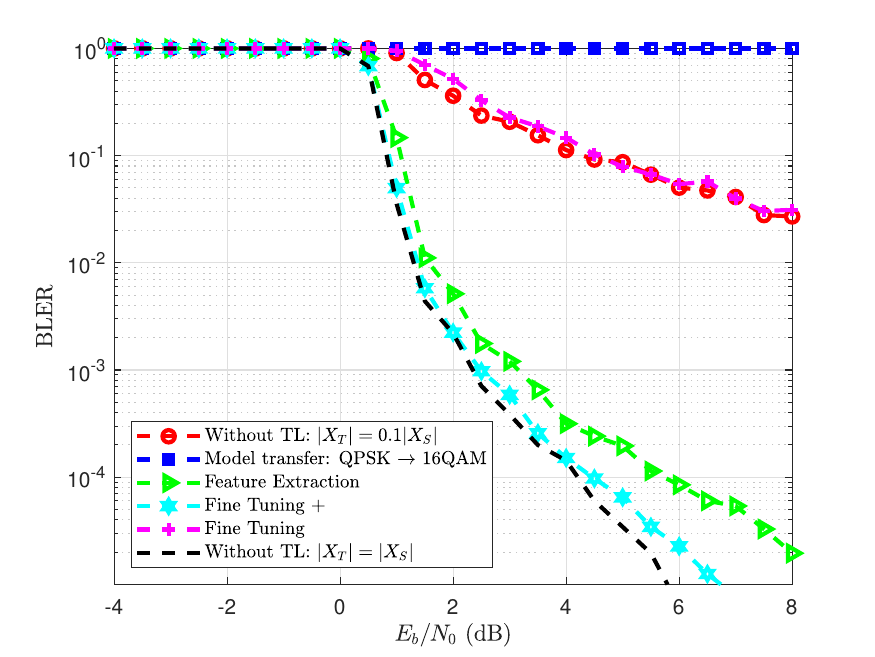}
 \caption{BLER Vs. $E_b/N_o$ for Transfer between QPSK to 16QAM}
 \label{fig_q}\vspace{-15pt}
 \end{figure}

 \begin{figure}[!t]
 \centering
 \includegraphics[width=1.0\linewidth]{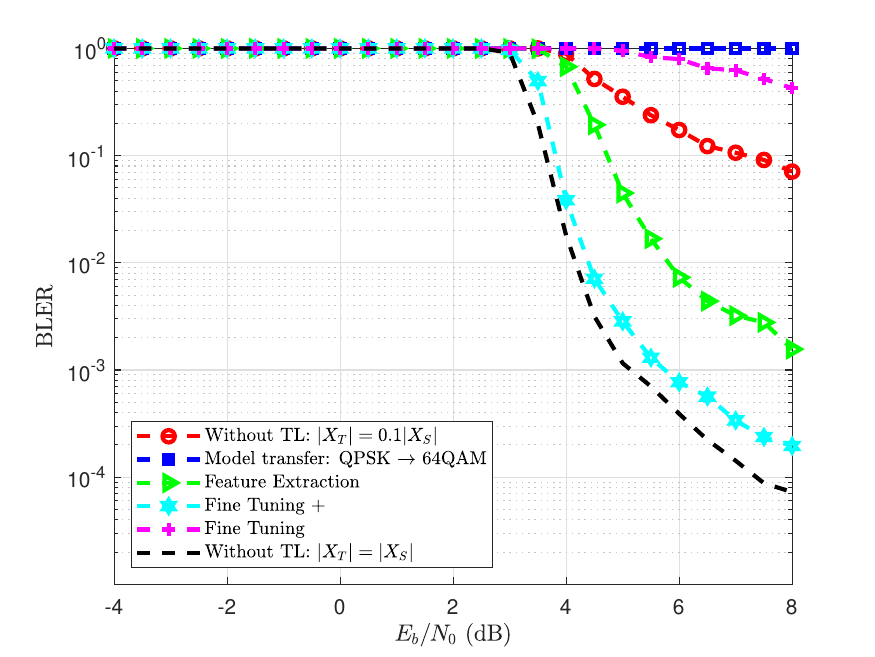}
 \caption{BLER Vs. $E_b/N_o$ for transfer between QPSK to 64QAM}
 \label{fig_q2}
 \end{figure}

 \subsection{Transfer Learning for Modulation Mismatch}This subsection considers transfer learning for modulation mismatch between the source and the target models. The source deep neural receiver is trained on symbols mapped to QPSK constellations. The resulting source model weights are transferred to a target model where fine-tuning or feature extraction is performed with a smaller dataset derived from \SI{16} or \SI{64} QAM constellations and thereafter tested with a test dataset derived from symbols mapped to \SI{16} or \SI{64} QAM respectively. That is, the dataset of the source model are received symbols which were mapped to QPSK constellations whereas at the target model, the dataset is made of received symbols either mapped to \SI{16} or \SI{64} QAM. The channel model used in both target and source models is the CDL-C channel model.

The plots in Figure~\ref{fig_q} show the performance of the transfer learning techniques and the aforementioned benchmarks. The very poor performance of the \textit{Model transfer QPSK $\rightarrow$ \SI{16}QAM} benchmark shows that  evaluating or testing a deep neural receiver on a dataset derived from a constellation different from the one it was trained on will degrade the receiver's decoding ability. This stems from the difference between the structure of these constellations. Furthermore, training the target model with a small target domain dataset (as \textit{Without TL :$|X_T|$=0.1$|X_S|$} shows) leads to a relatively poor performance because the dataset is insufficient to enable the deep neural receiver extract enough features to make predictions when presented with test data. As \textit{Fine Tuning} modifies all transferred/pretrained weights with a small target dataset, its performance shows that the pretrained weights should not only be used as initialisation points for retraining. The performances of  of both \textit{Feature extraction}  and \textit{Fine Tuning +} show that retaining some/all the transferred weights of the source model can improve the target deep neural receiver performance when exposed to signals derived from \SI{16}QAM modulation. This indicates that although QPSK and 16QAM have different constellation structures, the deep neural receiver is still able to extract useful features from the source dataset.

A similar trend is seen in Figure~\ref{fig_q2} where \textit{Fine Tuning +} reduces the gap between having no target dataset and having sufficient target dataset, for example at $1 \times 10^{-3}$ BLER, the gap between \textit{Model transfer QPSK $\rightarrow$ \SI{64}QAM} and \textit{Without TL :$|X_T|$=$|X_S|$} is cut down  to \SI{0.5}dB. The effect of the SINR range  used for evaluation is also seen in  Figure~\ref{fig_q2} as BLERs less than 1 are only observed after 3dB. With a higher modulation scheme (64 QAM), symbols are more tightly packed making reception more susceptible to noise.   Layer specific fine-tuning largely depends on source and target domains. Whereas for mismatched channels (as will be seen in a later subsection), fine-tuning the entire weights from the source model offers better receiver detection, for modulation mismatch, partial fine-tuning provides significant performance improvement over full fine-tuning.   

 \begin{figure}
  \centering\includegraphics[width=1.0\linewidth]{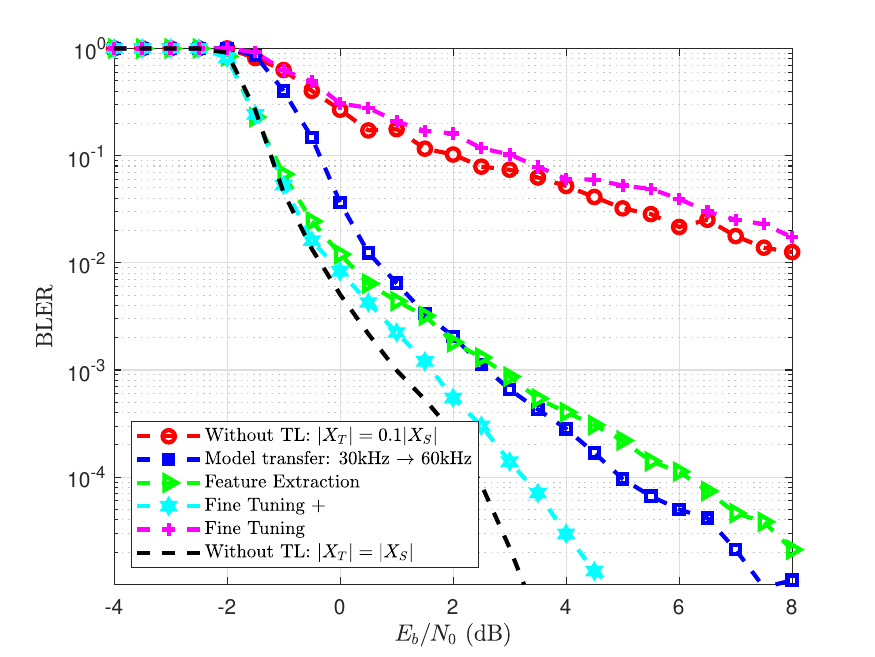}
 \caption{BLER versus $E_b/N_o$ for transfer between SCS 30kHz to 60kHz}
 \label{fig_scs}\vspace{-10pt}
 \end{figure}

 \begin{figure}
 \centering
   \includegraphics[width=1.0\linewidth]{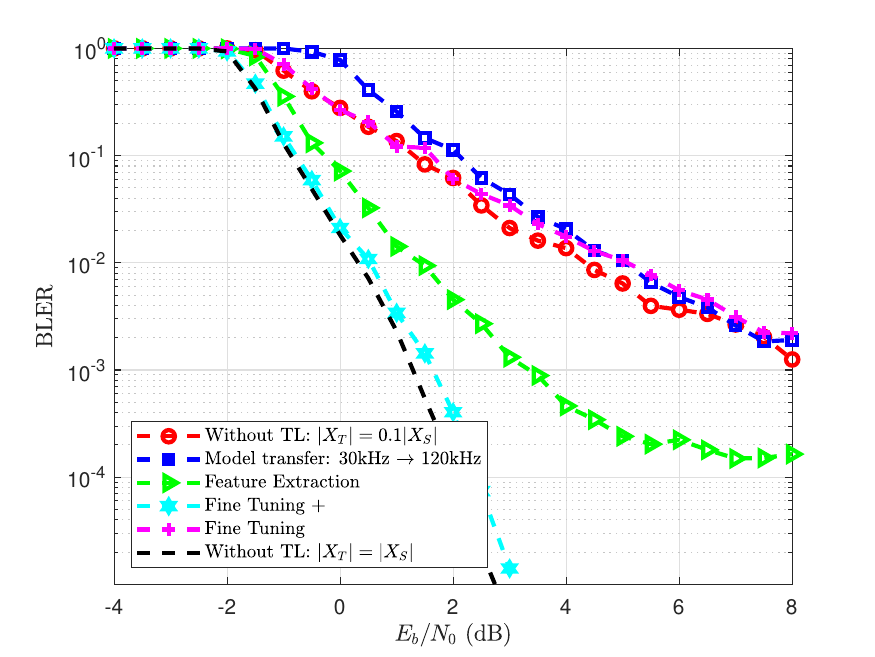}
 \caption{BLER Vs. $E_b/N_o$ for Transfer between SCS 30kHz to 120kHz}
 \label{fig_scs2}
 \end{figure}

\subsection{Transfer Learning for Sub-Carrier Spacing Mismatch}
In this subsection, we consider subcarrier spacing mismatch in which the datasets of the source and target domains are derived from different subcarrier spacings. The results show that increased mismatch between source and target models impacts the performance of the deep neural receiver.  Where the mismatch distance is small (i.e. transferring from \SI{30}kHz to \SI{60}kHz) as in Figure \ref{fig_scs}, the target model or receiver achieves reasonable performance with zero target domain data.  The performance of \textit{Fine Tuning} indicates that using the weights of the source model as only an initialization point for adapting the neural receiver to the target domain is not sufficient. On the other hand, retaining all the weights of the source model can be useful as shown in the performance of the \textit{Feature Extraction} approach. However a compromise between not fine tuning all the source model weights and not fine-tuning any of them tends to offer better adaptation performance.  Such trend is seen in Figure \ref{fig_scs2} where the \textit{Fine Tuning +} technique reduces the gap between not having  target domain dataset and training on sufficient target domain dataset to \SI{0.2}dB.   Varying subcarrier spacing can affect both the inter carrier and inter symbol interference of a considered communication system. Hence subcarrier spacing mismatch affects target deep neural receiver performance. 
For subcarrier spacing and modulation mismatch, the parameters of propagating environment are aligned and only the configuration of the transmitting device are dis-aligned. Partially finetuning the weights from the source model guarantees that a portion of the source (neural network) model parameters are unchanged leading to superior BLER performance which explains the superior performance of \textit{Fine Tuning +} in the modulation and subcarrier spacing mismatch cases.

\subsection{Transfer Learning for Channel Model Mismatch}
\begin{figure}
\centering
\includegraphics[width=1.0\linewidth]{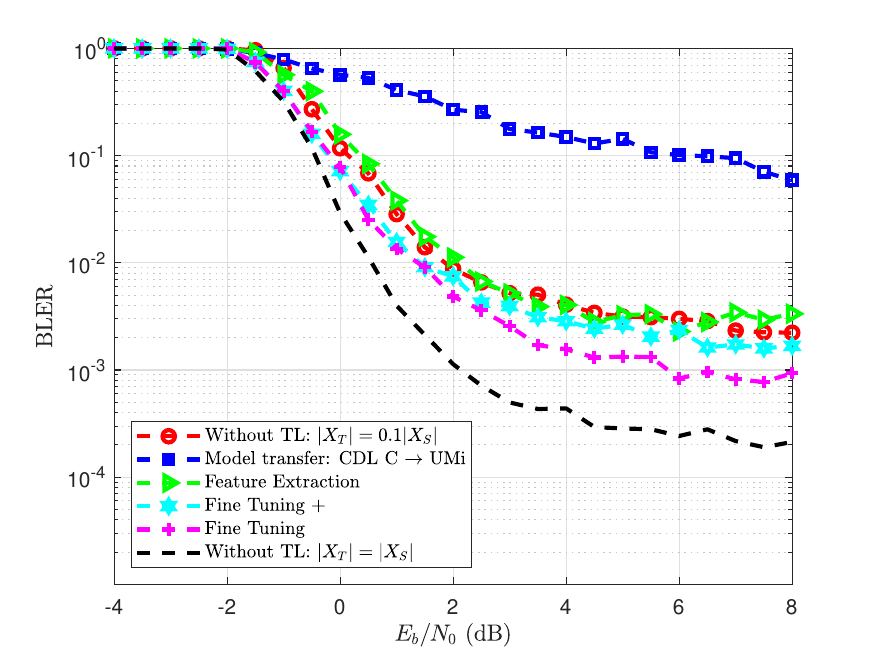}
\caption{BLER Vs. $E_b/N_o$ for Transfer between CDL C to 3GPP UMi}
\label{fig_6}\vspace{-15pt}
\end{figure}

\begin{figure}[!t]
\centering
\includegraphics[width=1.0\linewidth]{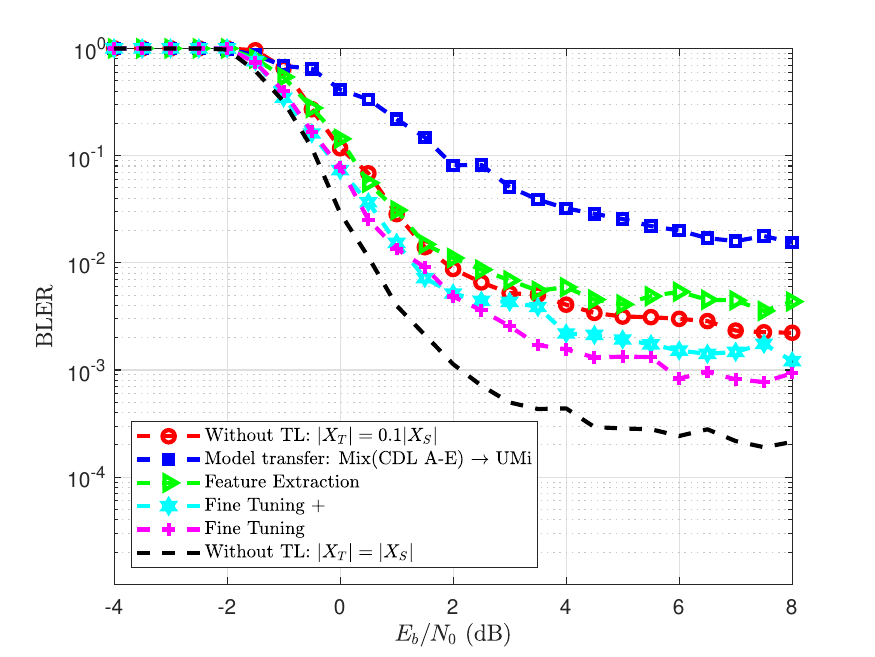}
\caption{BLER Vs. $E_b/N_o$ for Transfer between Mixed CDL (A-E)  to UMi}
\label{fig_7}\vspace{-10pt}
\end{figure}

\begin{figure}[!t]
\centering
\includegraphics[width=1.0\linewidth]{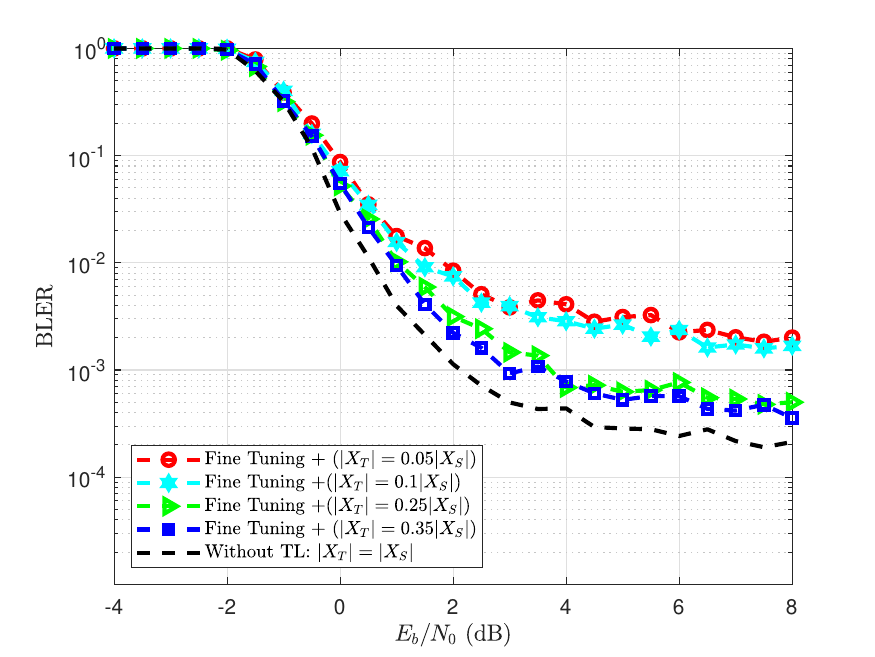}
\caption{BLER for Various $|X_T|$ }
\label{fig_9}\vspace{-10pt}
\end{figure}

 In Figure \ref{fig_6}, the performance of the aforementioned techniques are shown for CDL-C to 3GPP UMi channel model mismatch. The  \textit{Model transfer C $\rightarrow$ UMi} shows the poorest performance. This performance is due to having 0 target dataset to modify the weights transferred from the source deep neural receiver trained in CLD-C. Both fine-tuning techniques offer some performance improvement over training the deep neural receiver from scratch with a smaller target dataset (as \textit{Without TL: $|X_T|= 0.1|X_S|$} shows). However, \textit{Fine Tuning} offers slightly better performance than \textit{Fine Tuning +} to the tune of 
about \SI{0.5}dB at \SI{2e-2} BLER. This trend goes against the trend observed in the previous mismatch cases where \textit{Fine Tunig +} far outperformed \textit{Fine Tuning}. Hence for channel model mismatch, all transferred source weights are a good initialisation for adapting the deep neural receiver to a different channel model. 
Basically with target dataset size of $0.1|X_S|$, fine-tuning enhances the decoding ability of a  deep neural receiver  with transferred weights from a source deep neural receiver trained on dataset derived from CDL-C channel model such that when presented with dataset derived from 3GPP UMi, the SNR saving at \SI{1e-2} BLER is \SI{0.4} dB over training from scratch with sufficient target domain dataset. Propagating environment plays a critical role in wireless communication. With channel model mismatch, the parameters of the propagating environment in the source model and target model are not aligned, and so full finetuning of the weights from the source model offers better BLER performance. For subcarrier spacing and modulation mismatch, the parameters of propagating environment are aligned and only the configuration of the transmitting device are dis-aligned. Partially finetuning the weights from the source model guarantees that a portion of the source (neural network) model parameters are unchanged leading to superior BLER performance which explains the superior performance of \textit{Fine Tuning +} in the modulation and subcarrier spacing mismatch cases. It is needful to note that transferring from 3GPP UMi to CDL A-E does not need any target dataset at the target domain and so the results are not included here. \textit{Feature extraction}'s performance points to the limitation of modifying only the newly added layers in the considered case of channel model mismatch. 
Similar to the plots in Figure \ref{fig_6}, the plots in Figure \ref{fig_7} considers 3GPP UMi as the target domain. However in   Figure \ref{fig_7}, the source domain is a mixture of received signals derived from all the CDL channels (CDL A-E). The results show that such mixed training only improves the target neural receiver performance when there is 0 data at the target domain (\textit{Model transfer: A-E $\rightarrow$ UMi}). Such improvement is seen in the near \num{5}$\times$ reduction in BLER at an SNR of \SI{2}dB compared to the performance of \textit{Model transfer: A-E $\rightarrow$ UMi} in Figure \ref{fig_6}. The results also show that a mixture of CDL A-E channel models does not fully describe or approximate to an UMi channel model.
To show the effect of the amount of target dataset ($|X_S|$) available for fine tuning on the performance of the \textit{Fine tuning +} approach, various sizes of target dataset were considered in Figure \ref{fig_9} where again the target dataset consists of symbols derived from the 3GPP UMi channel model. 
The results show that \textit{Fine Tuning +} can nearly match the performance of \textit{Without TL:}$|X_T|=|X_S|$ but using a smaller target dataset. Thus a deep neural receiver having a dataset size that is only a fraction of the source dataset size can offer good decoding performance by modifying some source transferred weights using the target dataset. From the plots, having a target dataset of size which is 35\% of source dataset size can reduce the performance gap to nearly \SI{1}dB at $1 \times 10 ^{-3}$ BLER.
Concerning complexity, it is noteworthy that the number of trainable parameters in the target model of \textit{Fine Tuning} is \num{4858882} out of \num{4858882} total parameters, the number of trainable parameters of the target model in \textit{Feature Extraction} is \num{1214978} out of \num{6071554} total parameters whereas in \textit{Fine Tuning +}, the number of trainable parameters is \num{4852994} out of \num{6071554} total parameters. Therefore, the number of parameter updates during transfer learning is least in \textit{Feature Extraction} and highest in \textit{Fine Tuning}.

\section{Conclusion}
In this paper, we have evaluated the performance of some transfer learning techniques for mismatches in channel model, modulation schemes and subcarrier spacing using a deep neural receiver. Full finetuning showed superior performance in channel model mismatch whereas partial fine-tuning performed best in the other configuration mismatches. Overall partially modifying transferred layer weights using target data enhances deep neural receiver's adaptability to new environments. Future work will consider site specific channel models and other few shot learning techniques.

\bibliographystyle{IEEEtran}

\bibliography{tl_arxiv}

\balance

\end{document}